# The Heats of Reactions. Calorimetry and Van't-Hoff. 3

I. A. Stepanov

Latvian University, Rainis bulv. 19, Riga, LV-1586, Latvia

## Abstract

Earlier it has been supposed that the law of conservation of energy for chemical reactions has the following form:

$$dU = dQ - PdV + \sum_i \mu_i dN_i$$

In [1] it has been shown that for the biggest part of reactions it must have the following form:

$$dU = dQ + PdV + \sum_i \mu_i dN_i$$

In the present paper this result is confirmed by other experiments.

## 1. Introduction

For chemical processes the law of conservation of energy is written in the following form:

$$dU = dQ - PdV + \sum_i \mu_i dN_i \qquad (1)$$

where dQ is the heat of reaction, dU is the change in the internal energy, $\mu_i$ are chemical potentials and $dN_i$ are the changes in the number of moles.

In [1] it has been shown that the energy balance in the form of (1) for the biggest part of the chemical reactions is not true. For the biggest part of the chemical reactions the law of conservation of energy must have the following form:

$$dU=dQ+PdV+\sum_i \mu_i dN_i \qquad (2)$$

The Van't-Hoff equation is the following one:

$$\partial \ln K/\partial T = \Delta H^0/RT^2 \qquad (3)$$

where K is the reaction equilibrium constant and $\Delta H^0$ is the enthalpy. According to thermodynamics, the Van't-Hoff equation must give the same results as calorimetry because it is derived from the 1st and the 2nd law of thermodynamics without simplifications. However, there is a paradox: the heat of chemical reactions, that of dilution of liquids and that of other chemical processes measured by calorimetry and by the Van't-Hoff equation differ significantly [1]. The difference is far beyond the error limits. The reason is that in the derivation of the Van't-Hoff equation it is necessary to take into account the law of conservation in the form of (2), not of (1) [1].

If to derive the Van't-Hoff equation using (2), the result will be the following one:

$$\partial \ln K/\partial T = \Delta H^{0*}/RT^2 \qquad (4)$$

where $\Delta H^{0*}=\Delta Q^0+P\Delta V^0$.

In the present paper the following processes were investigated: $2Cu_2O=4Cu+O_2$, $2MgO=2Mg(g)+O_2$, $Mo_3Si+SiO_2(\alpha\text{-quartz})=3Mo+2SiO(g)$, $3Mo_5Si_3+4SiO_2(\alpha\text{-quartz})= 5Mo_3Si+8SiO(g)$, $5MoSi_2+7SiO_2(\alpha\text{-quartz})=Mo_5Si_3+14SiO(g)$, $2Al_2O_3=Al(g)+AlO(g)+Al_2O(g)+4O$, $Al(g)+O=AlO(g)$, $Al_2O(g)=Al(g)+AlO(g)$,

$2CdTe = 2Cd(g) + Te_2(g).$

## 2. Experiments

In [2, 4] the following reactions were considered:

$$2Cu_2O = 4Cu + O_2 \qquad (7)$$

$$2MgO = 2Mg(g) + O_2 \qquad (8)$$

The heat of these reactions is given in Table 1.

In [5] the following reactions were given:

$$Mo_3Si + SiO_2(\alpha\text{-quartz}) = 3Mo + 2SiO(g) \qquad (11)$$

$$\ln K = 0,524 - 803,83/RT, \qquad (1512\text{-}1663 \text{ K}) \qquad (12)$$

$$3Mo_5Si_3 + 4SiO_2(\alpha\text{-quartz}) = 5Mo_3Si + 8SiO(g) \qquad (13)$$

$$\ln K = 2,052 - 3064,7/RT, \qquad (1408\text{-}1674 \text{ K}) \qquad (14)$$

$$5MoSi_2 + 7SiO_2(\alpha\text{-quartz}) = Mo_5Si_3 + 14SiO(g) \qquad (15)$$

$$\ln K = 3,671 - 5145,6/RT \qquad (16)$$

The heat of these reactions is in Table 1.

In [10, 11, 13] the following reactions were given:

$$2Al_2O_3 = Al(g) + AlO(g) + Al_2O(g) + 4O \quad [10] \qquad (17)$$

$$Al(g) + O = AlO(g) \quad [11] \qquad (18)$$

$$Al_2O(g) = Al(g) + AlO(g) \quad [11] \qquad (19)$$

$$2CdTe = 2Cd(g) + Te_2(g) \quad [13] \qquad (20)$$

There heats are given in Table 1.

## 3. Conclusions

One sees that heats of chemical reactions obey the present theory, not the traditional one.

# Table 1

The heat of some chemical reactions measured by the Van't-Hoff equation and by calorimetry

| Reaction | T, K | $\Delta H^{*0}$ (kJ/mol) | $\Delta H^{*0} - P\Delta V^0$ (kJ/mol) | $\Delta Q$ (kJ/mol) |
|---|---|---|---|---|
| $2Cu_2O=4Cu+O_2$ [2] | 900 | 343,58 | 336,10 | 336,72 [3] |
| | 1000 | 343,58 | 335,27 | 335,38 [3] |
| | 1100 | 343,58 | 334,44 | 333,99 [3] |
| | 1200 | 343,58 | 333,61 | 332,57 [3] |
| | 1300 | 343,58 | 332,78 | 331,64 [3] |
| $2MgO=2Mg(g)+O_2$ [4] | 1900 | 1504,2 | 1456,9 | 1462,4 [3] |
| | 2000 | 1504,2 | 1454,3 | 1453,3 [3] |
| $Mo_3Si+SiO_2=3Mo+2SiO(g)$ [5] | 1600 | 803,83 | 777,24 | 776,37 [6, 7] |
| $3Mo_5Si_3+4SiO_2=5Mo_3Si+8SiO(g)$ [5] | 1200 | 3064,7 | 2984,9 | 2987,1 [6-8] |
| | 1300 | 3064,7 | 2978,3 | 2966,3 [6-8] |
| | 1400 | 3064,7 | 2971,6 | 2945,4 [6-8] |
| | 1500 | 3064,7 | 2965,0 | 2924,1 [6-8] |
| | 1600 | 3064,7 | 2958,3 | 2902,6 [6-8] |
| $5MoSi_2+7SiO_2=Mo_5Si_3+14SiO(g)$ [5] | 1400 | 5145,6 | 4982,7 | 4984,6 [6, 8, 9] |
| | 1500 | 5145,6 | 4971,1 | 4947,5 [6, 8, 9] |

| Reaction | T, K | $\Delta H^{*0}$ (kJ/mol) | $\Delta H^{*0}-P\Delta V^0$ (kJ/mol) | $\Delta Q$ (kJ/mol) |
|---|---|---|---|---|
| | 1600 | 5145,6 | 4959,5 | 4909,5 [6, 8, 9] |
| $2Al_2O_3=Al(g)+AlO(g)+Al_2O(g)+4O$ [10] | 2000 | 4684,0 | 4567,7 | 4526,1 [3, 12] |
| | 2100 | 4684,0 | 4561,8 | 4519,9 [3, 12] |
| | 2200 | 4684,0 | 4556,0 | 4513,5 [3, 12] |
| | 2300 | 4684,0 | 4550,2 | 4507,13 [3, 12] |
| $Al(g)+O=AlO(g)$ [11] | 2000 | -544,28 | -527,66 | -517,56 [3, 12] |
| | 2100 | -544,28 | -526,83 | -517,18 [3, 12] |
| | 2200 | -544,28 | -526,0 | -516,74 [3, 12] |
| | 2300 | -544,28 | -525,17 | -516,27 [3, 12] |
| | 2400 | -544,28 | -524,34 | -515,75 [3, 12] |
| | 2500 | -544,28 | -523,51 | -515,20 [3, 12] |
| | 2600 | -544,28 | -522,67 | -513,94 [3, 12] |
| | 2700 | -544,28 | -521,84 | -514,11 [3, 12] |
| | 2800 | -544,28 | -521,01 | -513,57 [3, 12] |
| $Al_2O(g)=Al(g)+AlO(g)$ [11] | 2000 | 564,8 | 548,19 | 540,86 [3, 12] |
| | 2100 | 564,8 | 547,35 | 541,28 [3, 12] |
| | 2200 | 564,8 | 546,52 | 541,76 [3, 12] |
| | 2300 | 564,8 | 545,69 | 542,28 [3, 12] |
| $2CdTe=2Cd(g)+Te_2(g)$ [13] | 1100 | 573,84 | 546,42 | 550,41 [12] |
| | 1200 | 573,84 | 543,92 | 544,04 [12] |

| | | 1300 | 573,84 | 541,43 | 537,33 [12] |

      The enthalpy for $SiO_2$, $SiO$ and $Mo$ is taken from [6], that for $Mo_3Si$ from [7], that for $Mo_5Si_3$ from [8], that for $MoSi_2$ is calculated as $\Delta_f H^0(300\ K) + \int C_P dT$, $\Delta_f H^0$ is from [9], $C_P$ is from [6].

      The enthalpy for $Al(g)$ is from [12], that for $AlO$, $Al_2O$, $Al_2O_3$, $O$ is from [3].